\documentclass[prd,aps,floats,preprint]{revtex4}

\usepackage{amsmath}
\usepackage{amssymb}
\usepackage{graphicx}
\usepackage{psfrag}

\def\be{\begin{equation}}
\def\ee{\end{equation}}
\def\bea{\begin{eqnarray}}
\def\eea{\end{eqnarray}}
\def\mpl{M_{\rm p}}

\begin{document}

\title{Cosmic Acceleration and Modified Gravity}

\author{Mark Trodden\footnote{trodden@physics.syr.edu}}

\affiliation{Department of Physics, Syracuse University \\
Syracuse, NY 13244-1130,
USA.}

\begin{abstract}
I briefly discuss some attempts to construct a consistent modification to General Relativity (GR) that might explain
the observed late-time acceleration of the universe and provide an alternative to dark energy. I mention the issues facing extensions to GR, illustrate these
with two specific examples, and discuss the resulting observational and theoretical obstacles. This article comprises an invited talk at the NASA workshop {\it From Quantum to Cosmos: Fundamental Physics Research in Space}.
\end{abstract}

\maketitle

\section{Introduction}
Approaches to the late-time acceleration of the universe may be divided into three broad classes. First, it is possible that there is some as yet undiscovered property of our existing model of gravity and matter that leads to acceleration at the current epoch. Into this category one might include the existence of a tiny cosmological constant and the possibility that the backreaction of cosmological perturbations might cause self-acceleration.

Second is the idea that there exists a new dynamical component to the cosmic energy budget. This possibility, with the new source of energy density modeled by a scalar field, is usually referred to as {\it dark energy}.

Finally, it may be that curvatures and length scales in the observable universe are only now reaching values at which an infrared modification of gravity can make itself apparent by driving self-acceleration~\cite{Dvali:2000hr,Deffayet:2000uy,Deffayet:2001pu,Freese:2002sq,Dvali:2003rk,Carroll:2003wy,Capozziello:2003tk,Vollick:2003aw,Flanagan:2003rb,Flanagan:2003iw,Vollick:2003ic,Soussa:2003re,Nojiri:2003ni,Carroll:2004de,Arkani-Hamed:2003uy,Gabadadze:2003ck,Moffat:2004nw,Clifton:2004st,Carroll:2006jn}. It is this possibility that I will briefly describe in this article, submitted to the proceedings of the NASA workshop {\it From Quantum to Cosmos: Fundamental Physics Research in Space}.

While I will mention a number of different approaches to modified gravity, I will concentrate on laying out the central challenges to constructing a successful modified gravity model and on illustrating them with a particular simple example. Detailed descriptions of some of the other possible ways to approach this problem can be found in the excellent contributions of Sean Carroll, Cedric Deffayet, Gia Dvali and John Moffat.

\section{The Challenge}
Although, within the context of General Relativity (GR), one doesn't think about it too often, the metric tensor contains, in principle, more degrees of freedom than the usual spin-2 {\it graviton} (See Sean Carroll's talk in these proceedings for a detailed discussion of this).

The reason why one doesn't hear of these degrees of freedom in GR is that the Einstein-Hilbert action is a very special choice, resulting in second-order equations of motion, which constrain away the scalars and the vectors, so that they are non-propagating. However, this is not the case if one departs from the Einstein-Hilbert form for the action. When using any modified action (and the usual variational principle) one inevitably frees up some of the additional degrees of freedom. In fact, this can be a good thing, in that the dynamics of these new degrees of freedom may be precisely what one needs to drive the accelerated expansion of the universe. However, there is often a price to pay.

The problems may be of several different kinds. First, there is the possibility that along with the desired deviations from GR on cosmological scales, one may also find similar deviations on solar system scales, at which GR is rather well-tested. Second is the possibility that the newly-activated degrees of freedom may be badly behaved in one way or another; either having the wrong sign kinetic terms (ghosts), and hence being unstable, or leading to superluminal propagation, which may lead to other problems.

These constraints are surprisingly restrictive when one tries to create viable modified gravity models yielding cosmic acceleration. In the next few sections I will describe several ways in which one might modify the action, and in each case provide an explicit, clean, and simple example of how cosmic acceleration emerges. However, I will also point out how the constraints I have mentioned rule out these simple examples, and mention how one must complicate the models to recover viable models.

\section{A Simple Model: $f(R)$ Gravity}
The simplest way one could think to modify GR is to replace the Einstein-Hilbert Lagrangian density by a general function $f(R)$ of the Ricci scalar 
$R$~\cite{Carroll:2003wy,Capozziello:2003tk,Barrow:rx,Barrow:xh,Barrow:hg,Magnano:bd,dobado,Schmidt:gb}.
\be
S=\frac{\mpl^2}{2}\int d^4 x\sqrt{-g}\, \left[R+f(R)\right] + \int d^4 x\sqrt{-g}\, {\cal L}_{\rm m}[\chi_i,g_{\mu\nu}] \ ,
\label{jordanaction}
\ee
where $\mpl\equiv (8\pi G)^{-1/2}$ is the (reduced) Planck mass and ${\cal L}_{\rm m}$ is the Lagrangian
density for the matter fields $\chi_i$.

Here, I have written the matter Lagrangian as ${\cal L}_{\rm m}[\chi_i,g_{\mu\nu}]$ to make explicit that in
this frame - the {\it Jordan} frame - matter falls along geodesics of the metric $g_{\mu\nu}$.

The equation of motion obtained by varying the action~(\ref{jordanaction}) is
\be
\left(1+f_R \right)R_{\mu\nu} - \frac{1}{2}g_{\mu\nu}\left(R+f\right)
+ \left(g_{\mu\nu}\Box -\nabla_\mu\nabla_\nu\right) f_R
=\frac{T_{\mu\nu}}{\mpl^2} \ ,
\label{jordaneom}
\ee
where I have defined $f_R\equiv \partial f/\partial R$.

Further, if the matter content is described as a perfect fluid, with energy-momentum tensor,
\begin{equation}
T_{\mu\nu}^m = (\rho_m + p_m)U_{\mu} U_{\nu} + p_m g_{\mu\nu}\ ,
\label{perfectfluid}
\end{equation} 
where $U^{\mu}$ is the fluid rest-frame four-velocity, $\rho_m$ is the energy density and $p_m$ is the pressure, then the fluid equation of motion is the usual continuity equation. 

When considering the background cosmological evolution of such models, I will take the metric to be of the flat Robertson-Walker form, $ds^2=-dt^2+a^2(t)d{\bf x}^2$. In this case, the usual Friedmann equation of GR is modified to become
\be
3H^2 -3f_R ({\dot H}+H^2)+\frac{1}{2}f+18f_{RR}H({\ddot H}+4H{\dot H})=\frac{\rho_m}{\mpl^2}
\label{jordanfriedmann}
\ee
and the continuity equation is
\be
{\dot \rho}_m +3H(\rho_m+p_m)=0 \ .
\label{jordancontinuity}
\ee
When supplied with an equation of state parameter $w$, the above equations are sufficient to solve for the
background cosmological behavior of the space-time and it's matter contents. For appropriate choices of the
function $f(R)$ it is possible to obtain late-time cosmic acceleration without the need for dark energy, 
although evading bounds from precision solar-system tests of gravity turns out to be a much trickier
matter, as we shall see.

While one can go ahead and analyze this theory in the Jordan frame, it is more convenient to perform a carefully-chosen conformal transformation on the metric, in order to render the gravitational action in the usual Einstein Hilbert form of GR. 

Following the description in~\cite{Magnano:1993bd}, consider the conformal 
transformation
\be
{\tilde g}_{\mu\nu} = \Omega(x^{\alpha}) g_{\mu\nu} \ ,
\label{conftrans}
\ee
and construct the function $r(\Omega)$ that satisfies
\be
1+f_R[r(\Omega)]=\Omega \ .
\ee
Defining a rescaled scalar field by $\Omega \equiv e^{\beta\phi}$, with 
$\beta\mpl\equiv\sqrt{2/3}$, the resulting action becomes
\bea
{\tilde S}=\frac{\mpl}{2}\int d^4 x\sqrt{-{\tilde g}}\, {\tilde R} &+&\int d^4 x\sqrt{-{\tilde g}}\, 
\left[-\frac{1}{2}{\tilde g}^{\mu\nu}(\partial_{\mu}\phi)\partial_{\nu}\phi -V(\phi)\right] \nonumber \\ 
&+&
\int d^4 x\sqrt{-{\tilde g}}\, e^{-2\beta\phi} {\cal L}_{\rm m}[\chi_i,e^{-\beta\phi}{\tilde g}_{\mu\nu}]\ ,
\label{einsteinaction}
\eea
where the potential $V(\phi)$ is determined entirely by the original form~(\ref{jordanaction}) 
of the action and is given by
\be
V(\phi)=\frac{e^{-2\beta\phi}}{2}\left\{e^{\beta\phi}r[\Omega(\phi)] - f(r[\Omega(\phi)]) \right\} \ .
\label{einsteinpotential}
\ee

The equations of motion in the Einstein frame are much more familiar than those in the Jordan frame, 
although there are some crucial subtleties. In particular, note that in general, test particles of the matter content $\chi_i$ do not freely fall along geodesics of the metric ${\tilde g}_{\mu\nu}$.

The equations of motion in this frame are those obtained by varying the action with respect to the metric ${\tilde g}_{\mu\nu}$
\be
{\tilde G}_{\mu\nu} = \frac{1}{\mpl^2}\left({\tilde T}_{\mu\nu} + T^{(\phi)}_{\mu\nu}\right) \ ,
\label{einsteineom}
\ee
with respect to the scalar field $\phi$
\be
{\tilde \Box}\phi = -\frac{dV}{d\phi}(\phi) \ ,
\label{scalareom}
\ee
and with respect to the matter fields $\chi_i$, described as a perfect fluid.

Once again, I will specialize to consider background cosmological evolution in this frame. The 
Einstein-frame line element can be written in familiar FRW form as
\be
ds^2 =-d{\tilde t}^2+{\tilde a}^2({\tilde t})d{\bf x}^2 \ ,
\label{einsteinFRWmetric}
\ee
where $d{\tilde t}\equiv\sqrt{\Omega}\, dt$ and ${\tilde a}(t)\equiv\sqrt{\Omega} \,a(t)$. The Einstein-frame matter energy-momentum tensor is then given by
\be
{\tilde T}_{\mu\nu}^m = ({\tilde \rho}_m + {\tilde p}_m){\tilde U}_{\mu} {\tilde U}_{\nu} + 
{\tilde p}_m {\tilde g}_{\mu\nu}\ ,
\label{einsteinperfectfluid}
\ee
where ${\tilde U}_{\mu}\equiv \sqrt{\Omega} \,U_{\mu}$, ${\tilde \rho}_m\equiv \rho_m/\Omega^2$ and 
${\tilde p}_m\equiv p_m/\Omega^2$.

\subsection{A Simple Example}
For definiteness and simplicity focus on the simplest
correction to the Einstein-Hilbert action; $f(R)=-\mu^4/R$, with $\mu$ a new parameter with units
of $[{\rm mass}]$.

The field equation for the metric is then 
\be 
\label{cdtteqn}
 \left(1+\frac{\mu^4}{R^2}\right)R_{\mu\nu} -
 \frac{1}{2}\left(1-\frac{\mu^4}{R^2}\right)Rg_{\mu\nu}
  + \mu^4\left[g_{\mu\nu} \nabla_{\alpha}\nabla^{\alpha}
 -\nabla_{(\mu}\nabla_{\nu)}\right]R^{-2}
  =\frac{T_{\mu\nu}^m}{\mpl^2} \ .
\ee

The constant-curvature vacuum solutions, for which $\nabla_{\mu}R=0$,
satisfy $R=\pm\sqrt{3}\mu^2$. Thus, there exists a constant-curvature vacuum solution which is de Sitter space. 
We will see that the de Sitter
solution is, in fact, unstable, albeit with a very long decay time
$\tau\sim \mu^{-1}$.

The time-time component of the field equations for this metric is
\be
\label{newfriedmann} 
3H^2 - \frac{\mu^4}{12({\dot
H}+2H^2)^3}\left(2H{\ddot H} + 15H^2{\dot
H}+2{\dot H}^2+6H^4\right) = \frac{\rho_M}{\mpl^2} \ .
\ee

As I have discussed, one may now transform to
the {\em Einstein frame}, where the gravitational Lagrangian takes
the Einstein-Hilbert form and the
additional degree of freedom appears as a fictitious scalar field $\phi$, with potential
\begin{equation} 
\label{cddtpotential} 
V(\phi)=\mu^2 \mpl^2
e^{-2\beta\phi}\sqrt{e^{\beta\phi}-1} \ , 
\end{equation} 
shown in the figure
below. 
\begin{figure}[htb] 
\label{potential} 
\includegraphics[width=8truecm, angle=270]{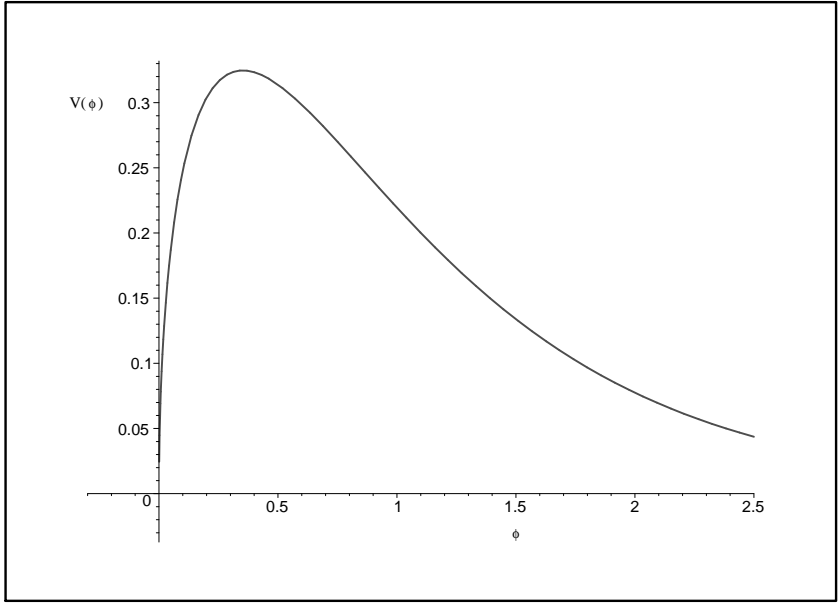}
\caption{The Einstein-frame potential $V(\phi)$} 
\end{figure}

Denoting with a tilde all quantities (except $\phi$) in the Einstein frame, 
the relevant Einstein-frame cosmological equations of motion are
\begin{equation}
 \label{conffriedmann}
 3{\tilde H}^2 = \frac{1}{\mpl^2}\left[\rho_\phi 
 + {\tilde\rho} \right] \ ,\\ 
\end{equation}
\begin{equation}
 \phi''+3\tilde H\phi'+\frac{dV}{d\phi}(\phi)- \frac{(1-3w)}{\sqrt{6}\mpl}
 {\tilde\rho}_M =0 \ , \label{confscalar}
\end{equation} 
where a prime denotes $d/d{\tilde t}$, and where 
\begin{equation}
 {\tilde\rho}_M = \frac{C}{{\tilde a}^{3(1+w)}}
 \exp\left[-\frac{(1-3w)}{\sqrt{6}}\frac{\phi}{\mpl} \right] \ , 
\end{equation} 
with $C$ a constant, and 
\begin{equation}
 \rho_\phi = \frac{1}{2}\phi'^2  + V(\phi) \ . 
\end{equation} 
Finally, note that the matter-frame Hubble parameter $H$ is related to that in the Einstein 
frame ${\tilde H}\equiv {\tilde a}'/{\tilde a}$ by 
\begin{equation}
 H = \sqrt{p} \left({\tilde H} -\frac{\phi'}{\mpl\sqrt{6}}\right) \ . 
\end{equation}

How about cosmological solutions in the Einstein frame? Ordinarily, Einstein gravity with a
scalar field with a minimum at $V=0$ would yield a Minkowski vacuum
state. However, here this is no longer true. Even though $V\rightarrow
0$ as $\phi\rightarrow 0$, this corresponds to a curvature singularity
and so is not a Minkowski vacuum. The other minimum of the 
potential, at $\phi\rightarrow \infty$, does not represent a solution.  

Focusing on vacuum solutions, i.e.,
$P_M=\rho_M = 0$, the beginning of the Universe corresponds to $R
\rightarrow \infty$ and $\phi \rightarrow 0$.  The initial conditions
we must specify are the initial values of $\phi$ and $\phi'$, denoted
as $\phi_i$ and ${\phi'}_i$. There are then three qualitatively distinct outcomes, depending
on the value of ${\phi'}_i$.

{\em 1.  Eternal de Sitter.}  There is a critical value of ${\phi'}_i
\equiv {\phi'}_C$ for which $\phi$ just reaches the maximum of the
potential $V(\phi)$ and comes to rest.  In this case the Universe
asymptotically evolves to a de~Sitter solution (ignoring spatial
perturbations).  As we
have discovered before (and is obvious in the Einstein frame), this
solution requires tuning and is unstable.

{\em 2.  Power-Law Acceleration.}  For ${\phi'}_i > {\phi'}_C$, the
field overshoots the maximum of $V(\phi )$. Soon thereafter, the
potential is well-approximated by $V(\phi) \simeq \mu^2\mpl^2
\exp(-\sqrt{3/2}\phi /\mpl)$, and the solution corresponds to $a(t)
\propto t^2$ in the matter frame. Thus, the Universe evolves to
late-time power-law inflation, with observational consequences similar
to dark energy with equation-of-state parameter $w_{\rm DE}=-2/3$.

{\em 3.  Future Singularity.}  For ${\phi'}_i < {\phi'}_C$, $\phi$
does not reach the maximum of its potential and rolls back down to
$\phi =0$.  This yields a future curvature singularity.

What about including matter? As can be seen from~(\ref{confscalar}), the major
difference here is that the equation-of-motion for $\phi$ in the
Einstein frame has a new term.  Furthermore, since the matter density
is much greater than $V \sim \mu^2\mpl^2$ for $t \ll 14\,$Gyr, this
term is very large and greatly affects the evolution of $\phi$.  The
exception is when the matter content is radiation alone ($w=1/3$), in
which case it decouples from the $\phi$ equation due to
conformal invariance. 

Despite this complication, it is possible to show that the three
possible cosmic futures identified in the vacuum case remain in the
presence of matter. 

Thus far, the dimensionful parameter $\mu$ is unspecified.
By choosing $\mu\sim 10^{-33}\,$eV, the corrections to the standard
cosmology only become important at the present epoch, explaining 
the observed acceleration of the
Universe without recourse to dark energy. 

Clearly the choice of correction to the gravitational action can be
generalized.  Terms of the form $-\mu^{2(n+1)}/R^n$, with $n>1$, lead
to similar late-time self acceleration, which can
easily accommodate current observational
bounds on the equation of
state parameter.

Now, as I mentioned in the introduction, any modification of the Einstein-Hilbert action must, of
course, be consistent with the classic solar system tests of gravity
theory, as well as numerous other astrophysical dynamical tests. 
We have chosen the coupling constant $\mu$ to be very small, but
we have also introduced a new light degree of freedom.  As shown by
Chiba~\cite{Chiba:2003ir}, the simple model above is equivalent to a Brans-Dicke
theory with $\omega=0$ in the approximation where the potential
was neglected, and would therefore be inconsistent with experiment~\cite{Bertotti:2003rm} (although see~\cite{Faraoni:2006hx,Capozziello:2005bu,Capozziello:2006jj} for suggestions that the conformally transformed theory may not be the correct way to analyze deviations from GR).

To construct a realistic $f(R)$ model requires a more complicated function, with more than one
adjustable parameter in order to fit the cosmological data and satisfy solar system bounds. Examples
can be found in~\cite{Nojiri:2003ni,Zhang:2005vt}.

\section{Extensions: Higher-Order Curvature Invariants}
It is natural to consider generalizing the action of~\cite{Carroll:2003wy} to include other curvature invariants.
There are, of course, any number of terms that one could consider, but for simplicity, focus on
those invariants of lowest mass dimension that are also parity-conserving
\begin{eqnarray}
P &\equiv & R_{\mu\nu}\,R^{\mu\nu} \ \nonumber \\
Q &\equiv & R_{\alpha\beta\gamma\delta}\,R^{\alpha\beta\gamma\delta} \ .
\end{eqnarray}

We consider actions of the form
\begin{equation}
S=\int d^4x \sqrt{-g}\,[R+f(R,P,Q)] +\int d^4 x\, \sqrt{-g}\,
{\cal L}_M \ ,
\label{genaction}
\end{equation}
where $f(R,P,Q)$ is a general function describing deviations from general relativity.

It is convenient to define
\begin{equation}
f_R\equiv\frac{\partial f}{\partial R}\ , \qquad
f_P\equiv\frac{\partial f}{\partial P}\ , \qquad 
f_Q\equiv\frac{\partial f}{\partial Q} \ ,
\end{equation}
in terms of which the equations of motion are
\begin{eqnarray}
R_{\mu\nu} &-& \tfrac12\,g_{\mu\nu}\,R-\tfrac12\,g_{\mu\nu}\,f \nonumber \\
&+& f_R\,R_{\mu\nu}+2f_P\,R^\alpha{}_\mu\,R_{\alpha\nu}
+2f_Q\,R_{\alpha\beta\gamma\mu}\,R^{\alpha\beta\gamma}{}_\nu\nonumber\\
&+& g_{\mu\nu}\,\Box f_R -\nabla_\mu\nabla_\nu f_R
-2\nabla_\alpha\nabla_\beta[f_P\,R^\alpha{}_{(\mu}{}\delta^\beta{}_{\nu)}]
+\Box(f_P\,R_{\mu\nu})\nonumber\\
&+& g_{\mu\nu}\,\nabla_\alpha\nabla_\beta(f_P\,R^{\alpha\beta})
-4\nabla_\alpha\nabla_\beta[f_Q\,R^\alpha{}_{(\mu\nu)}{}^\beta]
=8\pi G\,T_{\mu\nu}\ .\label{equaz}
\end{eqnarray}

It is straightforward to show that actions of the form~(\ref{genaction}) generically admit a maximally-symmetric solution: $R=$ a non-zero constant. However, an equally generic feature of such models is that this de Sitter solution is unstable. In the CDTT model the instability is to an accelerating power-law attractor. This is a possibility that we will also see in many of the more general models under consideration here.

Since we are interested in adding terms to the action that explicitly forbid flat space as a solution,
I will, in a similar way as in~\cite{Carroll:2003wy}, consider inverse powers of the above invariants
and, for simplicity, specialize to a class of actions with
\begin{equation}
f(R,P,Q)=-\frac{\mu^{4n+2}}{(aR^2+b P+c Q)^n} \ ,
\label{f_R}
\end{equation}
where $n$ is a positive integer (taken to be unity), $\mu$ has dimensions of mass and $a$, $b$ and $c$ are dimensionless constants. In fact, for general $n$ the qualitative features of the system are as for $n=1$\cite{Carroll:2004de}.

\subsection{Another Simple Example}
For the purposes of this short talk, I will focus on a specific example - actions containing modifications
involving only $P\equiv R_{\mu\nu}\,R^{\mu\nu}$, with the prototype being $f(P)=-m^6/P$, with $m$ a parameter with dimensions of mass.

It is easy to see that there is a constant curvature vacuum solution to this action given by $R_{\rm const}^{(P)} =\left(16\right)^{1/3}m^2$. However, we would like to investigate other cosmological solutions and analyze their stability.
 
From~(\ref{equaz}), with the flat cosmological ansatz, the analogue of the Friedmann equation becomes
\bea
3H^2&-&\frac{m^6}{8(3H^4+3H^2\dot H + \dot H^2)^3} 
\left[\dot H^4 + 11 H^2 \dot H^3 + 2 H \dot H^2\ddot H 
\right. \notag \\
&&\left.{}+33 H^4\dot H^2+
 30 H^6 \dot H +6 H^3 \dot H \ddot H +
 6 H^8 +  4 H^5\ddot H\right] =0 \ .
\label{riccieqn}
\eea

Asymptotic analysis of this equation (substituting in a power-law ansatz and taking the late-time limit)
yields two late-time attractors with powers $v_0=2-\sqrt{6}/2\simeq 0.77$ and $v_0=2+\sqrt{6}/2\simeq 3.22$.

However, in order to obtain a late-time accelerating solution ($p>1$), it is necessary to give accelerating initial conditions (${\ddot a}>0$), otherwise the system is in the basin of attraction of the non-accelerating attractor at $p\simeq 0.77$ (This type of behavior is generic in some other modified gravity theories~\cite{Easson:2005ax}).While I've given a simple example here, cosmologically viable models are described in~\cite{Mena:2005ta}.

What about the other constraints on these models? It has been shown~\cite{Navarro:2005gh} that solar system constraints, of the type I have described for $f(R)$ models, can be evaded by these more general 
models whenever the constant $c$ is nonzero. Roughly speaking, this is because the Schwarzschild 
solution, which governs the solar system, has vanishing $R$ and $P$, but non-vanishing $Q$.

More serious is the issue of ghosts and superluminal propagation. It has been shown~\cite{Chiba:2005nz,Navarro:2005da} that a necessary but not sufficient condition that the action be ghost-free is that $b=-4c$, so that there are no fourth derivatives in the linearised field equations. What remained was the possibility that the second derivatives might have the wrong signs, and also might allow superluminal propagation at some time in a particular cosmological background. It has recently been shown that in a FRW background with matter, the theories are ghost-free, but contain superluminally propagating scalar or tensor modes over a wide range of parameter space~\cite{DeFelice:2006pg,Calcagni:2006ye}. It is certainly necessary to be ghost-free. Whether the presence of superluminally propagating modes is a fatal blow to the theories remains to be seen.

\section{Conclusions}
Given the immense challenge posed by the accelerating universe, it is important to explore every option to explain the underlying physics. Modifying gravity may be one of the more radical proposals, but it is not one without precedence as an explanation for unusual physics. However, it is an approach that is tightly constrained both by observation and theoretical obstacles.

In the brief time and space allowed, I have tried to give a flavor of some attempts to modify GR to account for
cosmic acceleration without dark energy. I have focused on two of the directions in which I have been involved and have chosen to present simple examples of the models, which clearly demonstrate not only the
cosmological effects, but also how constraints from solar system tests and theoretical consistency apply.

There are a number of other proposals for modified gravity and, while I have had neither time nor space to devote to them here, others have discussed some of them in detail at this meeting.

There is much work ahead, with significant current effort, my own included, devoted to how one might distinguish between modified gravity, dark energy and a cosmological constant as competing explanations for cosmic acceleration.

\section*{Acknowledgments}
I would like to thank the organizers of the Q2C conference, and in particular Slava Turyshev, for their hard work and dedication in running such a stimulating meeting. I would also like to thank my many coauthors on the work discussed here for such enjoyable and productive collaborations, and for allowing me to reproduce parts of our work in this article. This work was 
supported in part by the NSF under grant PHY-0354990, by Research
Corporation, and by funds provided by Syracuse University


\end{document}